\documentclass[aps,prl,twocolumn,letterpaper,showpacs,superscriptaddress,floats]{revtex4}

\usepackage{graphics}

\begin{document}
\setlength{\parskip}{0mm}
\preprint{}

\title{Observed Effects of a Changing Step-Edge Density on Thin-Film Growth Dynamics}
\author{Aaron Fleet}
\affiliation{School of Applied and Engineering Physics, Cornell University, Ithaca, NY 14853}
\affiliation{Cornell Center for Materials Research, Cornell University, Ithaca, NY 14853}

\author{Darren Dale}
\affiliation{Department of Materials Science and Engineering, Cornell University, Ithaca, NY 14853}
\affiliation{Cornell Center for Materials Research, Cornell University, Ithaca, NY 14853}
\author{Y. Suzuki}
\affiliation{Department of Materials Science and Engineering, UC Berkeley, Berkeley, CA 94720}
\author{J. D. Brock}
\affiliation{School of Applied and Engineering Physics, Cornell University, Ithaca, NY 14853}
\affiliation{Cornell Center for Materials Research, Cornell University, Ithaca, NY 14853}

\date{\today}

\begin{abstract}
We grew SrTiO$_3$ on SrTiO$_3$ $[001]$ by pulsed laser deposition, while observing x-ray diffraction at the $(00\frac{1}{2})$ position.  The drop $\Delta I$ in the x-ray intensity following a laser pulse contains information about plume-surface interactions.  Kinematic theory predicts $\frac{\Delta I}{I}=-4\sigma(1-\sigma)$, so that $\frac{\Delta I}{I}$ depends only on the amount of deposited material $\sigma$.  In contrast, we observed experimentally that $|\frac{\Delta I}{I}|<4\sigma(1-\sigma)$, and that $\frac{\Delta I}{I}$ depends on the phase of x-ray growth oscillations.  The combined results suggest a fast smoothing mechanism that depends on surface step-edge density.
\end{abstract}


\maketitle

Particles of high kinetic energy can interact with a surface via a variety of exotic, atomistic processes.  A goal of modern surface science is to describe these interactions, which occur, for example, in thin-film growth by energetic particle beams, such as pulsed laser deposition (PLD).  Despite the widespread use of PLD to grow high quality crystalline films (especially functional complex-oxide materials \cite{imada_98,chrisey_94,lowndes_96}), a complete atomic-level description of the PLD process remains elusive.  In PLD, a megawatt laser pulse ablates the surface of a target material, creating a highly directed and extremely hot and dense ``plume'' of  ionic and neutral species \cite{willmott_00}.  A fundamental understanding of the interaction between the plume and the film surface would explain why PLD can produce atomically smooth films, rivalling thermal (molecular-beam epitaxy, chemical vapor deposition) and other energetic (sputter deposition, ion-beam assisted deposition) growth techniques.  The sensitivity of functional properties to film smoothness and crystallinity necessitates this understanding from a technological standpoint.  A complete model of PLD film growth may require inclusion of energetic processes not usually considered in standard models of growth. 

Conventional growth models typically assume that incident particles land at random positions on the substrate surface.  The particles then diffuse until they either (i) evaporate, (ii) attach to an existing step edge, or (iii) run into other particles and nucleate a new island \cite{BCF}.  Depending on the substrate temperature and specific energy barriers, film growth proceeds in one of several well-known modes: 3D, layer-by-layer, or step-flow  \cite{metev,neave,song}.  This treatment neglects the kinetic energy of the incident particles ($E_{K}\ll1$ eV for ``thermal'' sources), typically 10s of eV in the PLD plume \cite{lowndes_96}.  Other technologically important growth techniques such as sputter deposition and ion-beam assisted deposition also utilize particles with kinetic energies of 10s to 100s of eV, to enhance smooth growth.  Previous kinetic Monte Carlo-molecular dynamics simulations \cite{jacobsen_98, pomeroy_02}, and scanning-tunnelling microscopy studies of nucleation density  \cite{pomeroy_01, pomeroy_01b}, suggested that metal ions with energies near 20 eV can ``insert'' themselves into surface islands if they land near the island edges, during hyperthermal epitaxy. These insertion events occur during the initial impact and are complete well before particles come into thermal equilibrium with the substrate ($\tau_{thermal} \sim 1$ ps).  The efficacy of this insertion mechanism depends sensitively upon the step-edge density and thus upon the surface morphology.  In this letter, we report time-resolved x-ray scattering evidence of a qualitatively similar morphological effect during SrTiO$_3$ (STO) homoepitaxy via PLD.

Our PLD chamber is an integral component of a fully featured x-ray diffractometer that is permanently installed in the G3 hutch of the Cornell High Energy Synchrotron Source (CHESS).  By controlling the chamber oxygen pressure, $P_{O_{2}}$, substrate temperature, $T$, and laser repetition rate, $f$, we can select different growth modes, while monitoring film growth \emph{in-situ} with x-rays.  We grew STO films on STO $[001]$ substrates in two growth regimes: 1) $P_{O_{2}}=10$ mTorr, $T=540^{\circ}$C, $f=0.03$ Hz and 2) $P_{O_{2}}=0.01$ mTorr, $T=750^{\circ}$C, $f=0.1$ Hz.  We set the laser energy density at the single crystal STO target, and the target-substrate distance, to provide an average $\sigma\approx0.1\,$ monolayers/pulse.  To prepare a TiO$_{2}$-terminated surface, we etched the substrates in buffered NH$_{4}$F-HF  \cite{kawasaki,schrott}.  We annealed some substrates at $1000^{\circ}$C prior to growth to obtain a smooth surface \cite{jiang2,lippmaa}.

A key feature of x-ray scattering is that kinematic scattering theory accurately describes the scattered intensity and can directly test proposed growth models.  For the simple case of momentum transfer normal to the surface of a non-miscut crystal, one has, in homoepitaxy \cite{ocko_91},
\begin{equation}
I\propto\left | F \left ( q \right ) \right |^2 \,\left | \,\frac { 1 } { 1 - e^{i q d }}+\sum_n \, \theta_n \left ( t \right ) \, e^{- i q n d }\, \right |^2\; .
\label{eq:scatt_amp}
\end{equation}
Here $q$ is the scattering vector, $F \left ( q \right )$ is the scattering amplitude of a single layer, $\theta_n(t)$ is the time-dependent coverage of the $n^{th}$ layer, and $d$ is the layer spacing.  The first term represents the scattering from an ideally terminated single crystal.  The second term represents the scattering from the deposited film.  At the ``anti-Bragg'' $(00\frac{1}{2})$ position, x-rays scattered from adjacent layers interfere destructively, providing maximum sensitivity to single step height fluctuations.  By monitoring the anti-Bragg intensity,
one can monitor the layer coverage in real time and directly correlate the growth mode with the experimental conditions.  This technique has seen wide use in reflection high-energy electron diffraction  \cite{karl,lippmaa,blank,ohtomo}, and x-ray growth studies  \cite{woll_99,tischler,lee}.

\begin{figure}[b]
\resizebox{86mm}{!}{\includegraphics{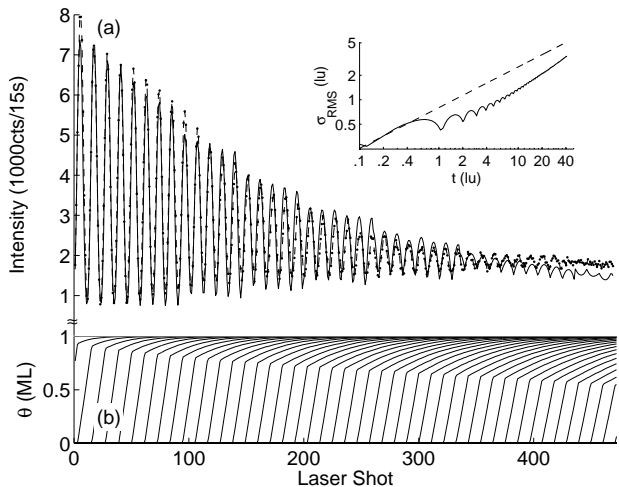}}
\caption{
(a) Anti-Bragg x-ray intensity oscillations during homoepitaxy of $\sim 42\,$ML of SrTiO$_{3}$; $P_{O_{2}}=10$ mTorr, $T=540^{\circ}$C, $f=0.03$ Hz.  The fit (solid line) to the data (dashed line) is generated using Eq.~(\ref{eq:scatt_amp}).  (b) Evolution of layer coverages $\theta_{n}$, determined by the fit.  Each line represents the growth of one ML.  Inset: Surface RMS roughness $\sigma_{RMS}$ versus film thickness $t$ during growth (both axes in lattice units).  Dashed line indicates random growth.
}
\label{growth}
\end{figure}
Under the conditions of regime 1, we monitored the anti-Bragg intensity during STO homoepitaxy [dots in Fig.~\ref{growth}(a)].  Each oscillation corresponds roughly to the completion of one monolayer. Each data point represents the scattered intensity integrated over a 15s window preceeding a laser pulse, up to, but not including, the fast change $\Delta I$ associated with the pulse.  The films initially grow in a quasi-2D layer-by-layer mode \cite{lippmaa,koster}.  After the first few layers, the growth takes on a 3D character.  The 3D growth is equivalent to roughening of the growing surface, which gradually decreases the oscillation maxima.  Additionally, a slight variation in the growth rate ($\sim 0.001$ monolayer/pulse) across the 1 cm sample causes the oscillation minima to increase during growth, as scattering from different regions of the surface becomes slightly out-of-phase. Accounting for these two factors, a fit to the data accurately describes the evolution of the anti-Bragg intensity over the entire $42$-layer deposition [solid line in Fig.~\ref{growth}(a)].  From the fit, we extracted the $\theta_{n}(t)$ [Fig.~\ref{growth}(b)].  Each $\theta_{n}(t)$ initially increases linearly as pulses of ablated material arrive at the surface.  As anticipated, new layers nucleate before completion of underlying layers, increasing the number of exposed layers as deposition proceeds. From the measured $\theta_{n}(t)$, we have calculated the time-evolution of the RMS roughness of the surface (inset to Fig.~\ref{growth}).  As expected, at late times, the roughness grows as $\sigma_{RMS} \sim t^{\beta},\beta=0.5$ \cite{barabasi}.

While this kind of measurement and analysis provides useful information on surface morphology and growth modes, much more information can be obtained by considering the pulsed nature of the growth.  In the impulse approximation, deposited material lands at random lattice sites on the (partially) exposed layers, instantaneously changing the layer coverages by $\Delta\theta_{n}=(\theta_{n}-\theta_{n+1})\sigma$. Calculating $\Delta I$ using $\Delta\theta_{n}$ and Eq.~(\ref{eq:scatt_amp}) leads directly to the elegantly simple result
\begin{equation}
\frac{\Delta I}{I}=-4\sigma(1-\sigma)\, ,
\label{eq:dI}
\end{equation}
where $I$ is the pre-pulse intensity.  Equation (\ref{eq:dI}) is the first significant new result of this work. Note that $\frac{\Delta I}{I}$ is a function of $\sigma$ only and is independent of both the surface morphology and the growth mode. It thus establishes a rigorous criterion for determining whether a system deviates from simple random deposition during pulsed homoepitaxy.

\begin{figure}[b]
\resizebox{86mm}{!}{\includegraphics{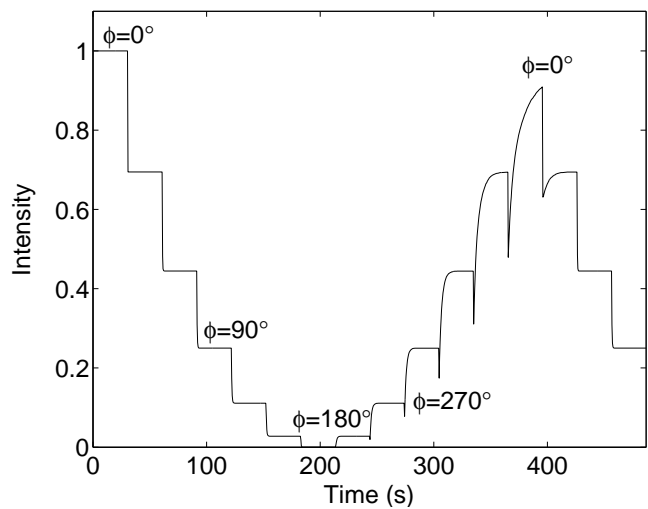}}
\caption{
Simulated quasi-2D growth of $1.2$ ML, at growth rate $\frac{1}{\sigma}=12$ pulses/ML.
}
\label{growthsim}
\end{figure}
To illustrate the significance of this result, we have simulated the anti-Bragg intensity during pulsed deposition of 1.2 monolayers of material onto a perfectly flat substrate.  Fig.~\ref{growthsim} illustrates both the instantaneous intensity drop $\Delta I$ associated with every pulse and the subsequent smoothing due to thermal relaxation.  The thermal smoothing is less obvious during the first half of the oscillation, because the overlayer coverage is small, so most deposited material lands on the substrate.  Thermal smoothing becomes visible when the overlayer exceeds half-coverage, because the majority of deposited material lands on the overlayer.  Note that $\Delta I$ is always negative.  Random deposition {\em always} roughens the surface.  The intensity oscillations associated with layer-by-layer growth require interlayer transport to smoothen the film.

To test the simple prediction of Equation (\ref{eq:dI}), we measured $\frac{\Delta I}{I}$ during STO homoepitaxy via PLD, under the conditions of growth regime 2, in which films grow in a near steady-state layer-by-layer mode \cite{ohtomo_02}.  To obtain the required time resolution, we used a multi-channel scaler (MCS) with 1 ms dwell time per channel, and 10000 channels per pass.  The MCS triggered the laser at the midpoint of each pass, so that we recorded 5s of detector intensity before and after each laser pulse.  During the 5 s prior to the laser pulse, interlayer diffusion has essentially concluded, so that the intensity is nearly constant.  For analysis, we normalized each MCS pass by the intensity $I$ integrated over the pre-pulse region.  Since $\frac{\Delta I}{I}$ should be independent of the surface morphology, we averaged all normalized MCS passes from one deposition.  We fit a straight line through the pre-pulse data, and an exponential through the post-pulse data \cite{karl}.  Evaluating the difference between the fits at $t=5$s gives $\frac{\Delta I}{I}=-0.108\pm 0.003$.
\begin{figure}[b]
\resizebox{86mm}{!}{\includegraphics{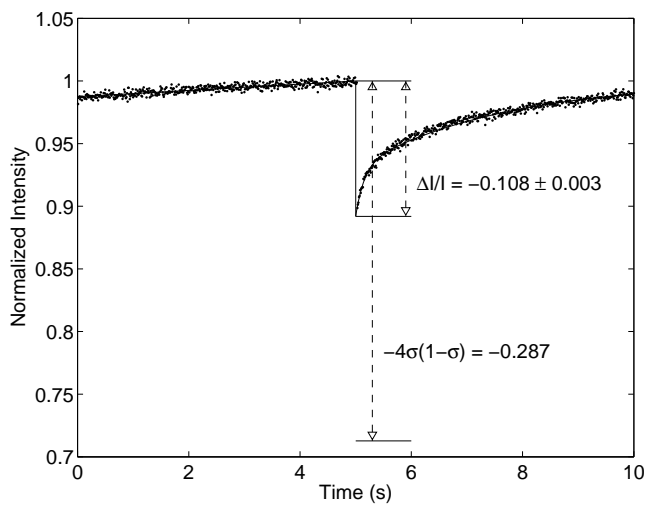}}
\caption{
Average of all normalized MCS passes for sample grown at $P_{O_{2}}=0.01$ mTorr, $T=750^{\circ}$C, $f=0.1$ Hz. Solid lines represent linear ($t<5$ s) and exponential ($t>5$ s) fits.
}
\label{mcs_all}
\end{figure}
From the growth oscillation data, we find $\sigma=0.078$ monolayers/pulse, and $-4\sigma(1-\sigma)=-0.287$, clearly indicating a discrepancy between
theory and experiment (Fig.~\ref{mcs_all}).  STO homoepitaxy via PLD is therefore more complex than the pulsed, random deposition implicit in the impulse approximation.  This is the second new result reported here.

The discrepancy widens dramatically when we consider the effect of the surface morphology.
\begin{figure}[b]
\resizebox{86mm}{!}{\includegraphics{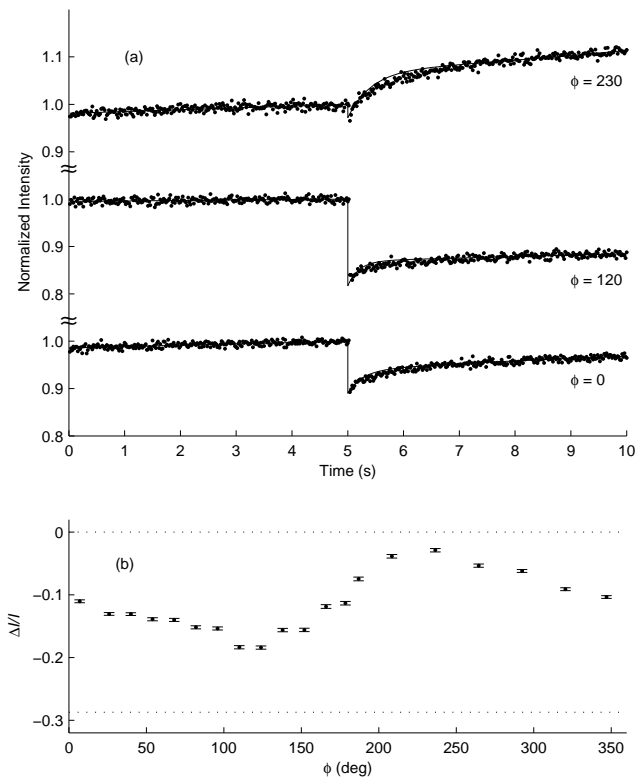}}
\caption{
(a) Average of normalized MCS data at three growth oscillation phases $\phi$.
(b) Measured $\frac{\Delta I}{I}$ at all experimental $\phi$.
Dotted lines represent theoretical upper and lower bounds.
}
\label{mcs_phases}
\end{figure}
Over the cycle of one intensity oscillation, the step-edge density varies as islands nucleate, grow and coalesce \cite{achutharaman}.  For purposes of discussion and analysis, we have assigned an angle $\phi$ to specify the phase of the intensity oscillations (Fig. \ref{growthsim}).  It is reasonable to assume that the density of step edges is lowest near intensity maxima ($\phi=0^{\circ}$), increases as islands nucleate and grow ($0^{\circ} < \phi \alt 180^{\circ}$), reaches a maximum ($\phi \approx 180^{\circ}$), and then decreases as the islands coalesce ($180^{\circ} \alt \phi < 360^{\circ}$).  Therefore, we interpret $\phi$ to be the phase of step-edge density oscillations.

The surface independence of $\frac{\Delta I}{I}$ in pulsed homoepitaxy equates with $\phi$-independence. To test this prediction, we averaged the normalized MCS passes at a particular $\phi$. A visual inspection of the data at three phases clearly reveals a morphological dependence to $\frac{\Delta I}{I}$ [Fig.~\ref{mcs_phases}(a)].  While the shapes of the intensity curves agree with qualitative predictions of the model [Fig.~\ref{growthsim}], the magnitude of $\frac{\Delta I}{I}$ clearly exhibits a phase dependence, with $\left | \frac{\Delta I}{I} \right | < 4\sigma(1-\sigma)$ at all $\phi$. This is the third new result reported here.  Fig.~\ref{mcs_phases}(b) illustrates quantitatively the dependence of $\frac{\Delta I}{I}$ on $\phi$.

The particular morphological dependence exhibited in Fig.~\ref{mcs_phases} is consistent with an energetic smoothing mechanism, the activation of which depends on step-edge density. The data ({\em e.g.}, Fig.~\ref{mcs_all}) demonstrate that thermal smoothing occurs on time-scales on the order of seconds.  More importantly, they reveal the existence of another smoothing mechanism, instantaneous relative to the 1ms binning times used here.  Interestingly, the magnitude of $\frac{\Delta I}{I}$ most closely approaches the theoretical value near $\phi\approx120^{\circ}$, and not at growth oscillation maxima ($\phi=0^{\circ}$), where one expects the lowest density of step edges.  However, the smallest step-edge density can occur at $\phi>0^{\circ}$ if small islands nucleate on top of a rapidly coalescing underlayer.  The maximum step-edge density would occur just prior to coalescence, which could coincide with the maximum deviation of $\frac{\Delta I}{I}$ ($\phi\approx230^{\circ}$).

In summary, the non-thermal smoothing mechanism is very fast, and its efficacy is correlated with the step-edge density.  Presumably, the physical nature of the mechanism is more complex than the insertion mechanism suggested for noble metal systems.  However, step edges may provide chemically favorable incorporation sites.  Highly kinetic plume species landing near these sites would rapidly overcome potential barriers to film incorporation.  An alternative explanation would be a $\phi-$dependent deposition rate, a possibility we are currently investigating.  While this interpretation could explain the variation of $\frac{\Delta I}{I}$ with $\phi$, the analysis of Fig.~\ref{mcs_all} nonetheless supports rapid smoothing.

This type of time-resolved x-ray measurement and analysis is not restricted to PLD but applies equally well to all types of pulsed growth techniques where the impulse approximation applies.  Thus, any deposition beam that can be chopped on time-scales short relative to the characteristic time-scale for thermal transport can be studied. Examples of potential systems include electrostatically switched sputter deposition, hyper-thermal metal-ion beams, and pulsed supersonic beams. One can generalize Eq.~(\ref{eq:scatt_amp}) to apply this analysis to heteroepitaxy.

This work is supported by the Cornell Center for Materials Research, under National Science Foundation (NSF) Grant No.~DMR-0079992.  This research used the G-line facilities at CHESS.  The construction of the G-line facility was supported by the NSF under Grant No.~DMR-9970838.  CHESS is supported by the NSF and the NIH/NIGMS under Grant No.~DMR-0225180.  The authors gratefully acknowledge the advice and assistance of Kee-Chul Chang in STO surface preparation.

\bibliography{sources}

\end{document}